\documentclass[12pt]{article}

\usepackage{amssymb}
\usepackage{amsmath}
\usepackage{graphicx} 
\usepackage{xcolor}
\usepackage{caption}
\usepackage{subfigure}
\usepackage{psfrag}
\usepackage{color} 
\usepackage{hyperref}
\usepackage[normalem]{ulem}  
\title{Stability of oscillations in the spatially extended May-Leonard model}
\begin{document} 
\maketitle
\begin{center}
Idan Sorin, Alexander Nepomnyashchy
\end{center}
\begin{center}
Department of Mathematics, Technion - Israel Institute of Technology,\\ Haifa 32000, Israel
\end{center}
\begin{center}
Vladimir Volpert
\end{center}
\begin{center}
Department of Engineering Sciences \& Applied Mathematics,\\ Northwestern University, Evanston, IL 60208, USA
\end{center}
\begin{abstract}
\noindent The May-Leonard model 
for three competing species, symmetric with respect to cyclic permutation of the variables  and extended by diffusive terms, is considered. Exact time-periodic solutions of the system have been found, and their stability with respect to spatially periodic disturbances is studied. The stability of solutions with respect to long-wave spatial modulations is revealed. A period-doubling instability breaking the spatial uniformity is found.
\end{abstract}
\section{Introduction}
\label{sec:1}
The spontaneous appearance of spatially uniform oscillations has been first observed by Belousov 
\cite{Bel} and 
Zhabotinskii \cite{Zha} in chemical kinetics systems. That phenomenon is the manifestation of the generic 
long-wave oscillatory instability (type {\bf III}$_o$ according to the classification of Cross and 
Hohenberg \cite{CrHo}). Kuramoto and Tsuzuki \cite{KuTs75} described the small-amplitude chemical oscillations using the complex Ginzburg-Landau equation \cite{ArKr}. 
\noindent The spatially uniform oscillations governed by the complex Ginzburg-Landau equation can be subject to the 
instability creating spontaneous spatial phase modulation, which is similar to the Benjamin-Feir 
instability of water waves. The small-amplitude long waves of phase modulation created by that 
instability 
are governed by the Kuramoto-Sivashinsky equation \cite{Nep, KuTs76}, which is a paradigmatic model of 
spatiotemporal chaos \cite{YaKu, BJPV}.
While the criterion of the modulational instability of small-amplitude oscillations is well known, less is 
known about the spatial instability of finite-amplitude nonlinear uniform oscillations. In \cite{ACR, 
Ris}, the case of a limit cycle close to a homoclinic trajectory has been considered. Two generic
instabilities have been revealed: (i) the long-wave phase-modulation instability corresponding to 
multipliers larger than 1 and (ii) the finite-wavenumber instability corresponding to 
multipliers smaller than -1 (period-doubling instability).

In the present paper, we consider oscillations governed by the May-Leonard system, which is a
paradigmatic model that can possess a robust attracting heteroclinic cycle. It has first been suggested as
an ecological model of a cyclic competition of species \cite{MaLe} and then obtained in \cite{BuHe} in the
description of the dynamics of patterns generated by the K\"{u}ppers-Lortz instability \cite{KuLo} of
convection rolls in a rotating liquid layer.
\noindent The May-Leonard system has an interesting feature: though the system is dissipative, for some values of 
parameters, there is a certain conservation law that leads to the simultaneous existence of a continuum of 
periodic solutions corresponding to uniform oscillations with different amplitudes. That allows us to follow  the stability properties in the whole interval of the conserved parameter change, from small-amplitude 
oscillations to those close to a heteroclinic cycle that includes three heteroclinic trajectories. In
Section \ref{sec:2}, we briefly describe the relevant properties of the May-Leonard system. Section \ref{sec:3} contains the
formulation of the problem. Section \ref{sec:4} is devoted to the analysis of the stability of uniform oscillations
with respect to long-wave disturbances. The absence of a long-wave modulational instability is revealed. In 
Section \ref{sec:5}, the results of a full linear stability analysis are described. A finite-wavenumber, period-doubling
instability is demonstrated. Section \ref{sec:6} contains the
conclusions. Appendix A contains the description of the fundamental matrix for the linearized problem with zero wavenumber. Appendix B presents details of the long-wave stability theory for small-amplitude oscillations.
\section{The May-Leonard system}
\label{sec:2}
 \subsection{General properties}
The May-Leonard model is a particular case of the competitive Lotka-Volterra system, 
\begin{equation}
\frac{dN_i(t)}{dt} =r_i N_i(t)\left[1-\sum_{j=1}^n \alpha_{ij}N_j (t)\right]
;\;N_i\geq 0;\;i=1,\ldots,n,
\label{eq:1}
\end{equation}
which describes the temporal evolution of the population that consists of $n$ species competing for common resources. Here
\( N_i(t) \) is the number of individuals of the \(i\)th  species at time \(t\), \(r_i\) is the intrinsic growth rate of the \(i\)th  species, and \( \alpha_{ij} \) are competition coefficients measuring the extent to which the \(j\)th species affects the growth rate of the \(i\)th species.
May and Leonard  \cite{MaLe} have analyzed the system of three competing species with \(r_1\)= \(r_2\)= \(r_3=1\) and the circulant  competition matrix \\ 
\[
\begin{bmatrix}
1 &\alpha & \beta \\
\beta & 1 & \alpha \\
\alpha & \beta & 1
\end{bmatrix}
\]
 with $0\leq\beta<1<\alpha$. 
 Later, we denote $N_1=u$, $N_2=v$ and $N_3=w$, thus the model is:
\begin{subequations}
\begin{equation}
   \dot{u}=u(1-u-\alpha v- \beta w)
   \label{eq2a}
\end{equation}
\begin{equation}
   \dot{v}=v(1-v-\alpha w- \beta u) 
   \label{eq2b}
\end{equation}
\begin{equation}
   \dot{w}=w(1-w-\alpha u- \beta v)
   \label{eq2c}
\end{equation}
\label{eq2}
\end{subequations}
 where the  dot denotes the temporal derivative. The dynamics is considered in the octant $u$, $v$, $w\geq 0$.

Let us briefly describe the dynamics of the system (\ref{eq2}). It has five equilibrium points: $(0,0,0)$, $(1,0,0)$, $(0,1,0)$, $(0,0,1)$, and ``the coexistence point"
\begin{equation}
    u=v=w=\frac{1}{1+\alpha+\beta}.
    \label{eq3}
\end{equation}
The coexistence point is stable, if $\alpha+\beta<2$, and oscillatory unstable, if $\alpha+\beta>2$.
The planes $u=0$, $v=0$ and $w=0$ are invariant manifolds. On the plane $w=0$, there exists the heteroclinic trajectory leading from $(1,0,0)$ to $(0,1,0)$. Similarly, there exist two more heteroclinic trajectories: from $(0,1,0)$ to $(0,0,1)$ on the plane $u=0$ and from $(0,0,1)$ to $(1,0,0)$ on the plane $v=0$. Thus, for any $\alpha>1$, $\beta<1$, there exists a robust heteroclinic cycle $(1,0,0)\to(0,1,0)\to(0,0,1)\to(1,0,0)$. The stability analysis shows that the heteroclinic cycle is attracting if $\alpha+\beta>2$, and repelling, if $\alpha+\beta<2$ \cite{MaLe}.
\noindent Thus, the coexistence point is an attractor as $\alpha+\beta<2$, and the heteroclinic cycle is an attractor as $\alpha+\beta>2$. The dynamics at $\alpha+\beta=2$ is quite nonstandard, and is discussed in the next subsection in more detail. \\
\subsection{The case \texorpdfstring{$\alpha + \beta = 2$}{alpha + beta = 2} }
\begin{figure}
\centering
\includegraphics[width=1.1\linewidth]{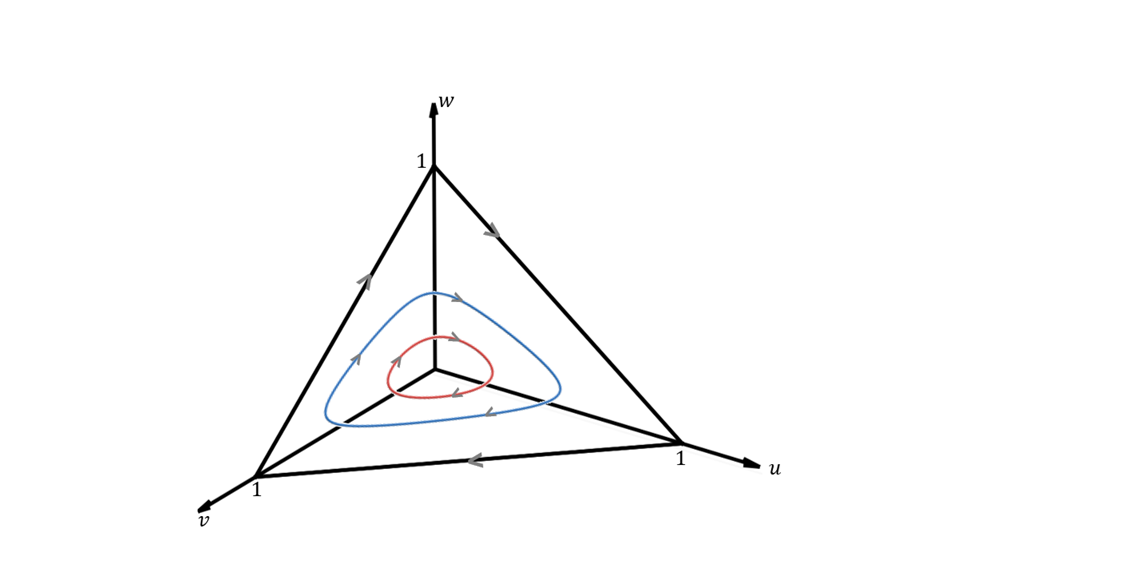}
\caption{\label{figure 1} Schematic of the system dynamics at $\alpha+\beta=2$. The triangle is the heteroclinic cycle and the  closed trajectories on the plane \(u+v+w=1\) correspond to periodic solutions with two different values of \(A\).}
\end{figure}

 Let us consider the temporal evolution of the sum $N=u+v+w$. In the case $\alpha+\beta=2$, we obtain the closed equation,
\begin{equation}
\dot{N}(t)=N(t)(1-N(t));
\label{eq4} 
\end{equation}
its solution is
\begin{equation}
N(t)=\frac{N(0)}{N(0)+[1-N(0)]e^{-t}}.
\label{eq5}
\end{equation}
 If $N(0)\neq 0$, then $N(t)\to 1$, as $t\to\infty$.
 Thus, the manifold $N=u+v+w=1$ is an attracting invariant manifold. It includes four fixed points,  namely the coexistence point $(1/3,1/3,1/3)$ and three saddle points, $(1,0,0)$, $(0,1,0)$ and $(0,0,1)$. 
\noindent Also, it contains heteroclinic trajectories connecting those saddle points, which form a triangle (see Figure \ref{figure 1}).  Note that the  heteroclinic trajectories can be found analytically. One of them is the segment of the line
\begin{equation}
w=0 ,\; u+v=1;
\label{eq6}
\end{equation}
the corresponding solution is:
\begin{equation}
u(t)= \frac{u(0)}{u(0)+(1-u(0))e^{t(\alpha -1)}} \ , \ v(t)= \frac{v(0)}{v(0)+(1-v(0))e^{-t(\alpha -1)}},\;-\infty<t<\infty
\label{eq7}
\end{equation}
with $v(0)=1-u(0)$. Two other heteroclinic trajectories and the corresponding solutions can be obtained from (\ref{eq5}), (\ref{eq6}) by a cyclic permutation of the variables $u$, $v$ and $w$.
\noindent Another important property of the dynamics at $\alpha+\beta=2$ is the existence of a conservation law. Let us rewrite the governing equations as

\begin{subequations}
 \begin{equation}   
  \frac{\dot{u}}{u}= \frac{d \ln(u)}{dt} =1-u-\alpha v- \beta w,
  \label{eq8a}
  \end{equation}
\begin{equation}  
  \frac{\dot{v}}{v}= \frac{d \ln(v)}{dt}=1-v-\alpha w- \beta u, 
  \label{eq8b}
  \end{equation}
\begin{equation}  
    \frac{\dot{w}}{w}= \frac{d \ln(w)}{dt}=1-w-\alpha u- \beta v.
    \label{eq8c}
\end{equation}
\label{eq8}
\end{subequations}
Summation of these  equations gives
$$\frac{d}{dt}\ln(uvw) = 3 - (1 + \alpha + \beta)(u + v + w) 
= 3(1 - N(t)) =$$
\begin{equation}
3\frac{N'(t)}{N(t)} 
= 3 \frac{d}{dt}\ln(N(t)) = \frac{d}{dt}\ln(N(t)^3),
\label{eq9}
\end{equation}
and therefore 
\begin{equation}
\frac{uvw}{(u+v+w)^3}=A=const.
\label{eq10}
\end{equation}
Thus, on the attracting manifold $u+v+w=1$, there exists a continuum of closed trajectories
\begin{equation}
uvw=A,\;0<A<\frac{1}{27},
\label{eq11}
\end{equation}
which correspond to periodic solutions. The value $A=1/27$ corresponds to the coexistence point $u=v=w=1/3$; the value $A=0$ is reached at the heteroclinic trajectories.
\subsection{ Analytical solutions}
\begin{figure}[h]
    \centering
    \includegraphics[width=0.7\textwidth]{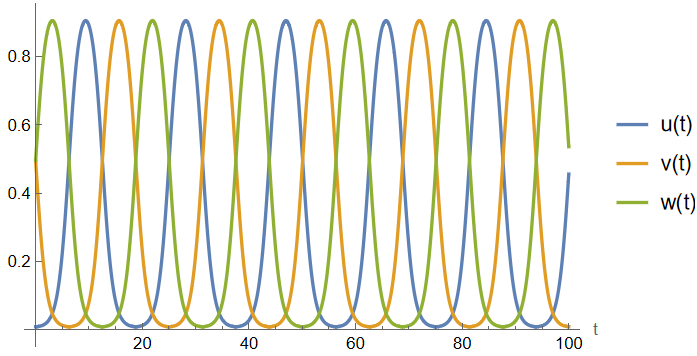}
    \caption{\label{figure2}The solution \((u(t),v(t),w(t))\) for \(A=2\cdot 10^{-3}\).}
\end{figure}

\begin{figure}[htbp]
\centering
\includegraphics[width=0.5\linewidth]{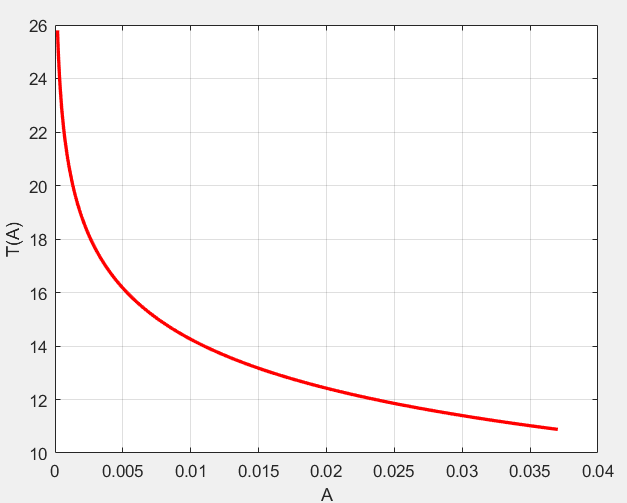}
\caption{\label{figure3} The  dependence of the period \(T\)} on \(A\).
\end{figure}

 Let us consider the case \(\alpha=2 \ , \ \beta=0:\)
\begin{subequations}
\label{eq12}
  \begin{equation}  
\dot{u}=u(1-u-2v)
\label{eq12a}
\end{equation}
\begin{equation} 
\dot{v}=v(1-v-2w)
\label{eq12b}
\end{equation}
\begin{equation}
\dot{w}=w(1-w-2u)
\label{eq12c}
\end{equation}
\end{subequations}
 On the attracting manifold $u+v+w=1$, the closed trajectory (\ref{eq11}) is described by the equation \( uv(1-u-v)=A\), hence
\begin{equation}
v=\frac{u-u^2 \pm \sqrt{(u-u^2)^2-4Au}}{2u}.
\label{eq13}
\end{equation}
We substitute (\ref{eq13}) 
into (\ref{eq12a}) and find that
\begin{equation}
\dot{u}=\pm \sqrt{P(u)},\;P(u)=(u-u^2)^2-4Au,\; 0<u<1.
\label{eq14}
\end{equation}
In (\ref{eq14}), one should take $`+'$ during the part of the period when $u$ grows, and $`-'$ when $u$ decreases. The solution of (\ref{eq14}) with the initial condition $u(0)=u_1$, $0<u_1<1$ is found as
\begin{equation}
t(u) = \frac{2}{\sqrt{u_2 (u_3 - u_1)}} \, \text{EllipticF} \left( \arcsin \left( \sqrt{ \frac{\frac{1}{u} - \frac{1}{u_1}}{\frac{1}{u_2} - \frac{1}{u_1}} } \right), \frac{u_3 (u_2 - u_1)}{u_2 (u_3 - u_1)} \right).
\label{eq15}
\end{equation}
The inverse  function is
\begin{equation}
u(t)= \frac{1}{\frac{1}{u_1}+(\frac{1}{u_2}-\frac{1}{u_1})\cdot sn^2\left[0.5t\sqrt{u_2(u_3-u_1)},\frac{u_3(u_2-u_1)}{u_2(u_3-u_1}\right]}
\label{eq16}
\end{equation}
 where \(u_1 < u_2 <u_3\) are the positive roots of the polynomial  $P(u)$ (the fourth root is zero), and
\(sn\) is the elliptic sine function.
Analyzing the properties of the polynomial $P(u)$ with \(0<A<1/27\), we conclude that \(u_1 , u_2<1 , u_3>1\),
so that \(u(t)\)  varies between \(u_1\) and \(u_2\), because the expression inside the square root should be positive, and the minimal and maximal values of \(u\)  correspond to \(\dot{u}=0\). Due to
the symmetry of the problem, the solutions $v(t)$, $w(t)$  are exactly the same as
\(u(t)\) but with a shift  by \(T/3\),
\begin{equation}
v(t)=u\left(t-\frac{T}{3}\right) \ , \  w(t)=u\left(t+\frac{T}{3}\right),
\label{eq17}
\end{equation}
 where  \(T\) is the period of (\ref{eq16}), which can be found  as
\(T=2t(u_2)\), i.e.,
\begin{equation}
T= \frac{4}{u_2(u_3-u_1)}K\left(\frac{u_3(u_2-u_1)}{u_2(u_3-u_1)}\right),
\label{eq18}
\end{equation}
 where $K$ is the complete elliptic integral of the first kind
(see Figures \ref{figure2} , \ref{figure3}).
The limits of $T$ at the boundaries of  the interval $(0, 1/27)$ are: 
\begin{equation}
\lim_{A\to 0} T = 2\pi K(1)=\infty, \;
\lim_{A\to 1/27} T = 2\pi \sqrt{3}.
\end{equation}
\section{Spatially extended May-Leonard system}
\label{sec:3}
 The subject of the present paper is the investigation of the stability of periodic solutions, which are described in the previous section, in the framework of the {\em spatially extended May-Leonard system},
\begin{subequations}
\label{eq20}
\begin{align}
u_t &= u(1 - u - 2v) + D_u u_{xx} \tag{20a}, \\
v_t &= v(1 - v - 2w) + D_v v_{xx} \tag{20b}, \\
w_t &= w(1 - w - 2u) + D_w w_{xx} \tag{20c};
\end{align}
\end{subequations}
$$0<t<\infty;\;-\infty<x<\infty$$
 (the subscripts $t$ and $x$ mean the partial derivatives with respect to the corresponding variables). The variables $u$, $v$ and $w$ depend now on both $x$ and $t$. The diffusion terms describe the spread of species  along the axis \(x\).
\noindent  Let us denote the spatially uniform, time-periodic solution of equation (\ref{eq20}) with a certain time period $T$, $2\pi\sqrt{3}<T<\infty$ as $(u_0(t),v_0(t),w_0(t))$. In order to investigate the linear stability of that solution, we impose a small periodic disturbance: 
\setcounter{equation}{20}
\begin{equation}
(u,v,w)=(u_0+\eta U(t)e^{ikx},v_0+\eta V(t)e^{ikx},w_0+\eta W(t)e^{ikx}).
\label{eq21}
\end{equation}
 Here \(0<\eta\ll 1\) is a small parameter, and \(k\) is the wavenumber of the perturbation.
 Substituting (\ref{eq21}) into (\ref{eq20})  and linearizing the obtained equations, we get the linear system of ODEs:
\begin{equation}
\begin{pmatrix}
\dot{U} \\
\dot{V} \\
\dot{W}
\end{pmatrix}
= - \begin{pmatrix}
D_u k^2 + 1 - 2w_0 & 2u_0 & 0 \\
0 & D_v k^2 + 1 - 2u_0 & 2v_0 \\
2w_0 & 0 & D_w k^2 + 1 - 2v_0
\end{pmatrix}
\begin{pmatrix}
U \\
V \\
W
\end{pmatrix}
\label{eq22}
\end{equation}
 (the property $u_0+v_0+w_0=1$ has been used).
 
Without loss of generality, let \(D_u\)=max \([D_u,D_v,D_w] \).  By rescaling of the wave-number, \(k \rightarrow k/\sqrt{D_u}\),  we get \( D_u=1, \ 0\leq D_v, D_w \leq 1\). Those relations will be used in the numerical calculations.

Let us discuss the general properties of system (\ref{eq22}). Because the matrix elements are periodic functions with the period \(T\),  the Floquet theorem \cite{Har}  states that the fundamental  matrix of  system (\ref{eq22}) for any $k$ can be presented as
\begin{equation}
\Phi(t;k)=P(t;k) e^{tB(k)}
\label{eq23}
\end{equation}
 where \(P(t;k)\) is a \(T\)-periodic matrix function and \(B(k)\) is a constant matrix.
The stability of the solution $(u_0(t),v_0(t),w_0(t))$ is determined by the eigenvalues $\lambda_i$, $i=1,2,3$ of matrix $B(k)$. 
If
\begin{equation}
{\mbox Re}\lambda_i<0 , i=1,2,3
\label{eq24}
\end{equation}
for all eigenvalues, the fundamental matrix \(\Phi(t;k)\) tends to zero as $t$ tends to \(\infty\), hence the
solution (\ref{eq16}), (\ref{eq17})  is stable.  
If \(Re{\lambda_j}>0\) for some \(j\in \{1,2,3\}\), that solution is unstable.
\section{Stability analysis for long-wave disturbances}
\label{sec:4}
We first analyze the stability with respect to long-wave disturbances, i.e.,  those with a small wavelength \(k\). We  apply the perturbation theory  in this case. First, we need to solve the system  (\ref{eq22}) at \(k=0\). 
\subsection{\texorpdfstring{The solution for $k=0$}{The solution for k=0}}
The analysis of system (\ref{eq22}) at $k=0$,
$$\dot{U}=-(1-2w_0)U-2u_0V,$$
\begin{equation}
\dot{V}=-(1-2u_0)V-2v_0W,
\dot{W}=-(1-2v_0)W-2w_0U,
\label{eq25}
\end{equation}

is given in Appendix A. It is shown there that one of the particular solutions of (\ref{eq25}) is just 
\begin{equation}
U_1(t)=\dot{u}_0 \ , \ V_1(t)=
\dot{v}_0 \ , \ W_1(t)=
\dot{w}_0.
\label{eq26}
\end{equation}
This solution, which corresponds to an infinitesimal shift along the closed orbit, is periodic with period $T$.
 
The second  particular solution, which corresponds to a shift across the orbit within the invariant plane $U+V+W=0$, i.e., a jump to an infinitesimally close orbit with different $A$, can be presented as:
\begin{equation}
U_2=U_{2p}(t)+c\dot{u_0}t \ , \ V_2=V_{2p}(t)+c\dot{v_0}t \ , \ W_2=W_{2p}(t)+c\dot{w_0}t,
\label{eq27}
\end{equation}
where $U_{2p}$, $V_{2p}$, $W_{2p}$ are periodic functions with the same period $T$ as the functions $u_0(t)$, $v_0(t)$, $w_0(t)$, and $c$ is a constant.  The linearly growing terms are caused by the dependence of the period $T$ on $A$.

The third particular solution, which describes the motion towards the attracting invariant plane $U+V+W=0$ along the surface (\ref{eq10}), is
\begin{equation}
U_3(t)=
U_{3p}(t)e^{-t},\;
V_3(t)
=V_{3p}(t)e^{-t},\;
W_3(t)
=W_{3p}(t)e^{-t},
\label{eq28}
\end{equation}
where $U_{3p}(t)$, $V_{3p}(t)$ and $W_{3p}(t)$ are periodic functions with period $T$. 

The fundamental matrix
\begin{equation}
\Phi(t)\equiv \Phi(t;0)=
\begin{pmatrix}
\dot{u_0} & U_{2p}+c\dot{u_0}t & U_{3p}e^{-t} \\
\dot{v_0} & V_{2p}+c\dot{v_0}t & V_{3p}e^{-t} \\
\dot{w_0} & W_{2p}+c\dot{w_0}t & W_{3p}e^{-t}
\end{pmatrix}
\label{eq29}
\end{equation}
can be presented in the form
\begin{equation}
\label{eq30}
\Phi(t) =
\begin{pmatrix}
\dot{u_0} & U_{2p} & U_{3p} \\
\dot{v_0} & V_{2p}& V_{3p} \\
\dot{w_0} & W_{2p} & W_{3p}
\end{pmatrix}
\begin{pmatrix}
1 & ct &0 \\
0 &1 & 0 \\
0 &0 & e^{-t}
\end{pmatrix}.
\end{equation}
Comparing with (\ref{eq23}), we find that
$$\Phi (t)=P(t)e^{tB}$$
with
\begin{equation}
\label{eq31}
P(t)=
\begin{pmatrix}
\dot{u_0} & U_{2p} & U_{3p} \\
\dot{v_0} & V_{2p}& V_{3p} \\
\dot{w_0} & W_{2p} & W_{3p}
\end{pmatrix} \ , \
e^{tB}=
\begin{pmatrix}
1 & ct &0 \\
0 &1 & 0 \\
0 &0 & e^{-t}
\end{pmatrix},
\end{equation}
and 
\begin{equation}
\label{eq32}
B=
\begin{pmatrix}
0 & c & 0 \\
0 & 0& 0 \\
0 & 0 &-1
\end{pmatrix}.
\end{equation}

\noindent Note that $\Phi(t)$ is the solution of the matrix equation,
\begin{equation}
\dot{\Phi}(t)=A(t)\Phi(t),
\label{eq33}
\end{equation}
where
\begin{equation}
A(t)=-\begin{pmatrix}
1-2w_0 & 2u_0 & 0 \\
0 & 1-2u_0 & 2v_0 \\
2w_0 & 0 & 1-2v_0
\end{pmatrix}.
\label{eq34}
\end{equation}

\noindent Due to the symmetry of equations
$$\dot{u}_0(t)=\dot{v}_0(t+T/3)=\dot{w}_0(t-T/3),$$
\begin{equation}
U_{2p}(t)=V_{2p}(t+T/3)=W_{2p}(t-T/3),\;
U_{3p}(t)=V_{3p}(t+T/3)=W_{3p}(t-T/3).
\label{eq35}
\end{equation}
Thus, the matrix $P(t)$ can be written in the following form:
\begin{equation}
P(t)=\begin {pmatrix} P_{11}(t) & P_{12}(t) & P_{13}(t)\\
P_{11}(t-T/3) & P_{12}(t-T/3) & P_{13}(t-T/3)\\
P_{11}(t+T/3) & P_{12}(t+T/3) & P_{13}(t+T/3) \end {pmatrix},   
\label{eq36} 
\end{equation}
and its determinant has the property
\begin{equation}
\mbox{det }P(t)=\mbox{det }P(t+T/3).
\label{eq37}
\end{equation}
Using (\ref{eq36}) and (\ref{eq37}), we can find a similar representation
for the inverse matrix $P^{-1}(t)$:
\begin{equation}
P^{-1}(t)=\begin {pmatrix}
(P^{-1})_{11}(t) & (P^{-1})_{11}(t-T/3) & (P^{-1})_{11}(t+T/3)\\
(P^{-1})_{21}(t) & (P^{-1})_{21}(t-T/3) & (P^{-1})_{21}(t+T/3)\\
(P^{-1})_{31}(t) & (P^{-1})_{31}(t-T/3) & (P^{-1})_{31}(t+T/3)
\end {pmatrix}.
\label{eq38} 
\end{equation}
\subsection{\texorpdfstring{The limit of small $k$}{The limit of small k}}
\subsubsection{Transformation of the problem}
We define
\begin{equation}
k^2=\epsilon \ , \ 0<\epsilon \ll 1
\label{eq39}
\end{equation}
and write (\ref{eq22}) in the form
\begin{equation}
\label{eq40}
\dot{y}=(A(t)+\epsilon C)y \ ,
\end{equation}
 where
\begin{equation}
y=\begin{pmatrix}
U \\
V \\
W
\end{pmatrix},
\label{eq41}
\end{equation}
$A(t)$ is determined by equation (\ref{eq34}), and
\begin{equation}
 C=\begin{pmatrix}
-D_u & 0 & 0 \\
0 & -D_v & 0 \\
0 & 0 &  -D_w
\end{pmatrix}.
\label{eq42}
\end{equation}
Applying the perturbation theory directly on equation (\ref{eq40}) is a  difficult task, since at the zeroth order we have a time-dependent matrix.
We suggest the following approach. Define:
\begin{equation}
y=P(t)x .
\label{eq43}
\end{equation}
As we have seen in the previous subsection, the solution of the equation \(\dot{\Phi}=A\Phi\) is
\(\Phi(t)=P(t)e^{tB}\),
so that
\begin{equation}
P=\Phi e^{-tB}
\label{eq44}
\end{equation}
Differentiating (\ref{eq44}) with respect to $t$, we find:
$$\dot{P}=\dot{\Phi}e^{-tB}-\Phi Be^{-tB}=A\Phi e^{-tB}-\Phi Be^{-tB}=APe^{tB}e^{-tB}-Pe^{tB}Be^{-tB}$$ 
therefore 
\begin{equation}
\dot{P}=AP-PB
\label{eq45}
\end{equation}
($B$ commutes with \(e^{tB}\)).
Thus,
\begin{equation}
\dot{y}=\dot{P}x+P\dot{x}=(AP-PB)x+P\dot{x}.
\label{eq46}
\end{equation}
Substituting (\ref{eq46}) into (\ref{eq40}), we obtain:
\begin{equation}
(AP-PB)x+P\dot{x}=(A+\epsilon C)Px.
\label{eq47}
\end{equation}
 Multiplying both sides of (\ref{eq47}) by $P^{-1}$ from the left, 
we get a perturbative problem
\begin{equation}
\dot{x}=(B+\epsilon D(t))x,\;D(t)=P^{-1}(t)CP(t),
\label{eq48}
\end{equation}
with a constant unperturbed matrix.

In what follows, we shall need to know the values of the components of the
matrix
$$\langle D\rangle=\frac{1}{T}\int_0^TD(t)dt.$$
\noindent First, let us consider the diagonal elements of that matrix. Using
expressions {(\ref{eq36})-(\ref{eq38})}, we find that
$$
D_{11}(t)=-D_u (P^{-1})_{11}(t) P_{11}(t)
 - D_v (P^{-1})_{11}(t-T/3) P_{11}(t-T/3)
 - D_w (P^{-1})_{11}(t+T/3) P_{11}(t+T/3)
$$
thus
\begin{align}
\langle D_{11} \rangle &= -\frac{D_u}{T} \int_0^T (P^{-1})_{11}(t) P_{11}(t) \, dt \nonumber \\
&\quad - \frac{D_v}{T} \int_0^T (P^{-1})_{11}(t - T/3) P_{11}(t - T/3) \, dt \nonumber \\
&\quad - \frac{D_w}{T} \int_0^T (P^{-1})_{11}(t + T/3) P_{11}(t + T/3) \, dt,
\label{eq49}
\end{align}
where the brackets $\langle...\rangle$ denote a time average over one oscillation period.
Because of the periodicity of the integrands, all the integrals in
(\ref{eq49}) are equal to each other. To find their value, let us
integrate the identity
$$(P^{-1}P)_{11}=(P^{-1})_{11}(t)P_{11}(t)+(P^{-1})_{11}(t-T/3)P_{11}(t-T/3)+
(P^{-1})_{11}(t+T/3)P_{11}(t+T/3)=1.$$
We find that each integral in (\ref{eq49}) is equal to $1/3$, hence
$$\langle D_{11}\rangle=-(D_u+D_v+D_w)/3.$$
The same is correct for $\langle D_{22}\rangle$ and
$\langle D_{33}\rangle$.

Similarly, using the relation
\(P^{-1}(t)P(t)=I\) and expressions (\ref{eq36})-(\ref{eq38}), we find that
\begin{equation}
 (P^{-1})_{j1}(t)P_{1k}(t) + (P^{-1})_{j1}(t-\frac{T}{3})P_{1k}(t-\frac{T}{3}) + (P^{-1})_{j1}(t+\frac{T}{3})P_{1k}(t+\frac{T}{3}) = \delta_{jk}, 
\label{eq50}
\end{equation}
therefore, all the non-diagonal elements of
$\langle D\rangle$ are equal to zero. Thus,
\begin{equation}
\langle D_{ij}\rangle=-\frac{1}{3}(D_u+D_v+D_w)\delta_{ij}.
\label{eq51}  
\end{equation}
\noindent Following the Floquet approach, we search the solutions of (\ref{eq48}) in the form
\begin{equation}
x(t)=z(t)e^{\lambda t}
\label{eq52}
\end{equation}
where 
$$z(t)=\begin{pmatrix}
z_1(t) \\
z_2(t) \\
z_3(t)
\end{pmatrix}$$ is a $T$-periodic vector function and \(\lambda\) is the eigenvalue.
We substitute (\ref{eq52})  into (\ref{eq48}) and get:
\begin{equation}
\dot{z}+\lambda z=Bz+\epsilon D(t)z;\;z(t+T)=z(t).
\label{eq53}
\end{equation}
Since the problem is three-dimensional, 
there are three  eigenvalues
\(\lambda\).
The stability of the solution  $(u_0(t),v_0(t),w_0(t))$ is determined by \(\lambda\): if \(Re \lambda<0\) \(x(t) \rightarrow 0\) and \(y(t) \rightarrow 0\) as \(t \rightarrow \infty\), so the solution is stable. If  \(Re \lambda>0\) at least for one of the eigenvalues, the solution is unstable.
\noindent First, let us consider the unperturbed problem
\begin{equation}
\dot{z}+\lambda z=Bz;\;z(t+T)=z(t).
\label{eq54}
\end{equation}
Obviously, the periodicity condition is satisfied only for time-independent vectors, $\dot{z}=0$, hence $\lambda$, $z$ are just the eigenvalues and eigenvectors of the matrix $B$ respectively. The eigenvalues of $B$ are $0$, $0$, $-1$.  The eigenvector corresponding to $\lambda_0=-1<0$ is
$$z_0=\begin{pmatrix}
0 \\
0 \\
1 \end{pmatrix}.$$ 
The corresponding perturbed solution is not interesting from the point of view of stability. Only one eigenvector,
\begin{equation}
\label{eq55}
z_{1,2}=\begin{pmatrix}
1 \\
0 \\
0 \end{pmatrix}
\end{equation}
corresponds to the double eigenvalue $\lambda_{1,2}=0$. The splitting of this eigenvalue by the perturbation is of the major interest from the point of view of stability.
\subsubsection{\texorpdfstring{Expansion in powers of $\epsilon^{1/2}$}{Expansion in powers of epsilon (1/2)}}
Because the matrix $B$ has a double eigenvalue $\lambda_{1,2}=0$ and only
one corresponding eigenvector, one can expect that the appropriate asymptotic expansions are power series in $\epsilon^{1/2}$ \cite{Hin}:
\begin{equation}
\lambda=\lambda^{(0)} +\epsilon^{1/2} \lambda ^{(1/2)} +... \ , \ z_i(t)=z_i^{(0)}+\epsilon^{1/2} z_i^{(1/2)}+...
\label{eq56}
\end{equation}
with $\lambda^{(0)}=0$. Taking into account expression (\ref{eq32}) for $B$, we obtain:
$$\left(\frac{d}{dt}+\epsilon^{1/2}\lambda^{(1/2)}+\ldots\right)
(z_1^{(0)}+\epsilon^{1/2}z_1^{(1/2)}+\ldots)=
c (z_2^{(0)}+\epsilon^{1/2}z_2^{(1/2)}+\ldots)+$$
$$\epsilon(D_{11}z_1^{(0)}+D_{12}z_2^{(0)}+D_{13}z_3)+\ldots,$$
$$\left(\frac{d}{dt}+\epsilon^{1/2}\lambda^{(1/2)}+\ldots\right)
(z_2^{(0)}+\epsilon^{1/2}z_2^{(1/2)}+\ldots)=
\epsilon(D_{21}z_1^{(0)}+D_{22}z_2^{(0)}+D_{23}z_3)+\ldots,$$
$$\left(\frac{d}{dt}+\epsilon^{1/2}\lambda^{(1/2)}+\ldots\right)
(z_3^{(0)}+\epsilon^{1/2}z_3^{(1/2)}+\ldots)=
-(z_3^{(0)}+\epsilon^{1/2}z_3^{(1/2)}+\ldots)+$$
$$\epsilon(D_{31}z_1^{(0)}+D_{32}z_2^{(0)}+D_{33}z_3)+\ldots.$$
\noindent At the zeroth order we have:
$$\dot{z}_1^{(0)}=cz_2^{(0)},\;\dot{z}_2^{(0)}=0,\;
\dot{z}_3{(0)}=-z_3{(0)}.$$
Thus, $z_2^{(0)}=z_2^{(0)}(0)$,
$z_1^{(0)}=z_1^{(0)}(0)+cz_2^{(0)}(0)t$,
$z_3^{(0)}=z_3^{(0)}(0)e^{-t}$. The periodicity condition gives
$z_2^{(0)}=0$, $z_3^{(0)}=0$. Below we choose $z_1^{(0)}=1$.
At the order $\epsilon^{1/2}$ we get:
$$\dot{z}_1^{(1/2)}+\lambda^{(1/2)}=cz_2^{(1/2)},$$
$$\dot{z}_2^{(1/2)}=0,\;\dot{z}_3{(1/2)}=-z_3{(1/2)}.$$
We find that $z_2^{(1/2)}(t)=z_2^{(1/2)}(0)$, and $z_1^{(1/2)}$ can
be periodic only if
\begin{equation}
\lambda^{(1/2)}=cz_2^{(1/2)}(0).
\label{eq57}
\end{equation}
Actually, $z_1^{(1/2)}(t)=z_1^{(1/2)}(0)$.
At the order $\epsilon$ we get:
$$\dot{z}_2^{(1)}+\lambda^{(1/2)}z_2^{(1/2)}=D_{21}(t).$$
Taking into account (\ref{eq57}), we find that the condition of
periodicity of $z_2^{(1)}$, $z_2^{(1)}(0)=z_2^{(1)}(T)$, is
$$[\lambda^{(1/2)}]^2\frac{T}{c}=\int_0^TD_{21}(t)dt.$$
Using (\ref{eq51}), we find that $\lambda^{(1/2)}=0$,
and the expansion in powers of $\epsilon^{1/2}$ is redundant: due
to the special symmetry properties of the perturbation matrix $D$,
it is sufficient to expand the eigenvalue and the solution in
powers of $\epsilon$.
\subsubsection{\texorpdfstring{Expansion in powers of $\epsilon$}{Expansion in powers of epsilon}}
 Now we expand \(z_i(t)\) and       \(\lambda\) to power series
\begin{equation}
\lambda=\lambda^{(0)} +\epsilon \lambda ^{(1)} +... \ , \ z_i(t)=z_i^{(0)}+\epsilon z_i^{(1)}+...
\label{eq58}
\end{equation}
with $\lambda^{(0)}=0$ and obtain three
equations:
\begin{equation}
\label{eq59}
\left(\frac{d}{dt}+\epsilon \lambda^{(1)}+...\right)(z_1^{(0)}+\epsilon z_1^{(1)}+...)=c(z_2^{(0)}+\epsilon z_2^{(1)}+...)+\epsilon \left(\sum_{j=1}^3 D_{1j}z_j^{(0)} +\ldots\right),
\end{equation}
\begin{equation}
\label{eq60}
\left(\frac{d}{dt}+\epsilon \lambda^{(1)}+...\right)(z_2^{(0)}+\epsilon z_2^{(1)}+...)=\epsilon \left(\sum_{k=1}^3 D_{2j}z_j^{(0)}+\ldots\right),   
\end{equation}
\begin{equation}
\label{eq61}
\left(\frac{d}{dt}+\epsilon \lambda^{(1)}+...\right)(z_3^{(0)}+\epsilon z_3^{(1)}+...)=-(z_3^{(0)}+\epsilon z_3^{(1)}+...)+\epsilon \left(\sum_{j=1}^3 D_{3j}z_j^{(0)}+\ldots\right).     
\end{equation}
\noindent The zeroth order solution is the same as before, hence
$$z_1^{(0)}=1,\;z_2^{(0)}=z_3^{(0)}=0.$$

\noindent At the order $O(\epsilon)$ we get:
\begin{equation}
\dot{z}_1^{(1)}+\lambda^{(1)}=cz_2^{(1)}+D_{11},
\label{eq62}
\end{equation}
\begin{equation}
\dot{z}_2^{(1)}=D_{21},
\label{eq63}
\end{equation}
\begin{equation}
\dot{z}_3^{(1)}=-z_3^{(1)}+D_{31}.
\label{eq64}
\end{equation}
\noindent Equation (\ref{eq63}) is solvable in the class of periodic
functions, because $\langle D_{21}\rangle =0$. The solution is:
\begin{equation}
z_2^{(1)}(t)=z_2^{(1)}(0)+\int_0^tD_{21}(\tau)d\tau.
\label{eq65}
\end{equation}
\noindent Below we denote
$$\langle f\rangle\equiv\frac{1}{T}\int_0^Tf(t)dt,\;\tilde{f}(t)=f(t)-\langle f\rangle$$
for any $T$-periodic function.

\noindent The periodicity condition gives
\begin{equation}
-\lambda^{(1)}+c\langle z_2^{(1)}\rangle+
\langle D_{11}\rangle=0. 
\label{eq66}
\end{equation}

\noindent If condition (\ref{eq66}) is satisfied, then
$${\dot{z}_1^{(1)}=c(z_2^{(1)}-\langle z_2^{(1)}\rangle)+D_{11}-\langle D_{11}\rangle} = c\tilde{z}_2^{(1)}(t)+\tilde{D}_{11},$$
 
\noindent hence,
\begin{equation}
z_1^{(1)}(t)=z_1^{(1)}(0)+
c\int_0^t\tilde{z}_2^{(1)}(\tau)d\tau+
\int_0^t\tilde{D}_{11}d\tau.
\label{eq67}
\end{equation}

\noindent The solution of (\ref{eq64}) is
\begin{equation}
z_3^{(1)}(t)=e^{-t}\left[z_3^{(1)}(0)+\int_0^tD_{31}(\tau)e^{\tau}d\tau\right].
\label{eq68}
\end{equation}  
The periodicity condition $z_3^{(1)}(T)=z_3^{(1)}(0)$ gives
\begin{equation}
z_3^{(1)}(0)=\frac{e^{-T}\int_0^TD_{31}(\tau)e^{\tau}d\tau}{\left(1-e^{-T}\right)}.
\label{eq69}
\end{equation}

\noindent At the order $\epsilon^2$,
\begin{equation}
\dot{z}_2^{(2)}+\lambda^{(1)}z_2^{(1)}=D_{21}z_1^{(1)}+D_{22}z_2^{(1)}+
D_{23}z_3^{(1)}.
\label{eq70}  
\end{equation}
The solvability condition for (\ref{eq70}) in the class of periodic functions is
\begin{equation}
\lambda^{(1)}\langle z_2^{(1)}\rangle=\langle D_{21}z_1^{(1)}\rangle +
\langle D_{22}z_2^{(1)}\rangle +\langle D_{23}z_3^{(1)}\rangle.
\label{eq71}  
\end{equation}
From (\ref{eq66}) we get
\begin{equation}
\langle z_2^{(1)}\rangle=\frac{1}{c}(\lambda^{(1)}-\langle D_{11}\rangle).
\label{eq72}
\end{equation}
By definition,
\begin{equation}
D_{22}=\langle D_{22}\rangle+\tilde{D}_{22}.
\label{eq73}
\end{equation}
Substituting (\ref{eq72}) and (\ref{eq73}) 
into (\ref{eq71}) and taking into account (\ref{eq51}), 
we obtain:
\begin{equation}
\frac{(\lambda^{(1)}-\langle D_{11}\rangle)
(\lambda^{(1)}-\langle D_{22}\rangle)}{c}=\langle
D_{21}\tilde{z}_1^{(1)}\rangle + \langle
\tilde{D}_{22}\tilde{z}_2^{(1)}\rangle +\langle D_{23}\tilde{z}_3^{(1)}\rangle,
\label{eq74}
\end{equation}
where
$$\langle D_{11}\rangle=\langle D_{22}\rangle=-\frac{1}{3}(D_u+D_v+D_w).$$
We find that
\begin{equation}
\lambda_{\pm}^{(1)}=-\frac{1}{3}(D_u+D_v+D_w)\pm\sqrt{R},
\label{eq75}
\end{equation}
where
\begin{equation}
R=c(\langle D_{21}\tilde{z}_1^{(1)}\rangle + \langle
\tilde{D}_{22}\tilde{z}_2^{(1)}\rangle +\langle D_{23}\tilde{z}_3^{(1)}\rangle).
\label{eq76}
\end{equation}
\noindent If $R>0$, $\lambda_{\pm}^{(1)}$ are real. If $R<0$, $\lambda_{\pm}^{(1)}$ 
are complex, and
$$\mbox{Re}\lambda_{\pm}^{(1)}=-\frac{1}{3}(D_u+D_v+D_w)<0,$$
hence in the latter case the spatially homogeneous oscillations are stable
with respect to long-wave modulations.
\subsubsection{Functional dependence of the eigenvalue splitting parameter on the
diffusion coefficients}
Let us define
\begin{equation}
F_{jk}(t)=(P^{-1})_{j1}(t)P_{1k}(t)
\label{eq77}
\end{equation}
and rewrite (\ref{eq50}) as
\begin{equation}
\label{eq78}
 F_{jk}(t) + F_{jk}\left(t-\frac{T}{3}\right) + F_{jk}\left(t+\frac{T}{3}\right) = \delta_{jk}.
\end{equation}

\noindent Let us expand the periodic  functions $F_{jk}(t)$ into Fourier series:
\begin{equation}
\label{eq79}
F_{jk}  = \sum_{n=-\infty}^{\infty} \hat{F}_{jk,n}\exp\left(i\frac{2\pi n}{T} t\right),\;F_{jk,-n}=F_{jk,n}^*.   
\end{equation}
Then equation (\ref{eq78}) gives:
\begin{equation}
\label{eq80}
\sum_{n=-\infty}^{\infty} \hat{F}_{jk,n}\exp\left(i\frac{2\pi n}{T}t\right)\left[1+\exp\left(-i\frac{2\pi n}{3}\right)+\exp\left(i\frac{2\pi n}{3}\right)\right]=\delta_{jk}. 
\end{equation}
The expression
$$1+\exp\left(-i\frac{2\pi n}{3}\right)+\exp\left(i\frac{2\pi n}{3}\right)=1+2\cos\left(\frac{2\pi n}{3}\right)$$
is equal to 3 if \(n (\mbox{mod 3})=0\) and $0$ otherwise. Therefore,
$$\hat{F}_{jk,0}=\frac{1}{3}\delta_{jk};\;\hat{F}_{jk,3m}=0,\;m=1,2,\ldots.$$
Because
$$D_{jk}(t)=-\left[D_uF_{jk}(t)+D_vF_{jk}\left(t-\frac{T}{3}\right)+
D_wF_{jk}\left(t+\frac{T}{3}\right)\right],$$
that expression can be written as
\begin{equation}
D_{jk}(t)=-\sum_n\hat{F}_{jk,n}\exp\left(i\frac{2\pi
n}{T}t\right)\left[D_u+D_v\exp\left(-i\frac{2\pi n}{3}\right)+D_w
\exp\left(i\frac{2\pi n}{3}\right)\right].
\label{eq81}
\end{equation}
Then we find:
$$\tilde{z}_2^{(1)}(t)=\int_0^tD_{21}(\tau)d\tau-\langle\int_0^tD_{21}(\tau)d\tau\rangle=$$
\begin{equation}
\sum_{n\neq 0}\frac{iT}{2\pi n}\hat{F}_{21,n}\exp\left(i\frac{2\pi n}{T}t\right)\left[D_u+D_v\exp\left(-i\frac{2\pi n}{3}\right)+D_w
\exp\left(i\frac{2\pi n}{3}\right)\right],
\label{eq82}
\end{equation}
$$\tilde{z}_1^{(1)}(t)=\int_0^t(c\tilde{z}_2^{(1)}(\tau)+\tilde{D}_{11}(\tau))d\tau-\langle \int_0^t(c\tilde{z}_2^{(1)}(\tau)+\tilde{D}_{11}(\tau))d\tau\rangle=$$
\begin{equation}
\sum_{n\neq 0}\left[c\left(\frac{T}{2\pi n}\right)^2\hat{F}_{21,n}+\frac{iT}{2\pi n}\hat{F}_{11,n}\right]\exp\left(i\frac{2\pi n}{T}t\right)\left[D_u+D_v\exp\left(-i\frac{2\pi n}{3}\right)+D_w
\exp\left(i\frac{2\pi n}{3}\right)\right],
\label{eq83}
\end{equation}
and
\begin{equation}
\tilde{z}_3^{(1)}(t)=\sum_{n\neq 0}\frac{\hat{F}_{31,n}}{1+2n\pi i/T}\exp\left(i\frac{2\pi n}{T}t\right)\left[D_u+D_v\exp\left(-i\frac{2\pi n}{3}\right)+D_w
\exp\left(i\frac{2\pi n}{3}\right)\right].
\label{eq84}
\end{equation}
\noindent Using (\ref{eq81}) and (\ref{eq82}), we find:
$$\langle D_{21}\tilde{z}_1^{(1)}\rangle=\sum_{n\neq 0}\left[c\left(\frac{T}{2\pi n}\right)^2|\hat{F}_{21,n}|^2-i\frac{T}{2\pi n}\hat{F}_{21,n}\hat{F}_{11,n}^{*}\right]\times$$
$$\left[D_u+D_v\exp\left(-i\frac{2\pi n}{3}\right)+D_w
\exp\left(i\frac{2\pi n}{3}\right)\right]\left[D_u+D_v\exp\left(i\frac{2\pi n}{3}\right)+D_w
\exp\left(-i\frac{2\pi n}{3}\right)\right].$$
The expression 
$$\left[D_u+D_v\exp\left(-i\frac{2\pi n}{3}\right)+D_w
\exp\left(i\frac{2\pi n}{3}\right)\right]\left[D_u+D_v\exp\left(i\frac{2\pi n}{3}\right)+D_w
\exp\left(-i\frac{2\pi n}{3}\right)\right]=$$
$$D_u^2+D_v^2+D_w^2+2D_uD_v\cos\left(\frac{2\pi n}{3}\right)+2D_uD_w\cos\left(\frac{2\pi n}{3}\right)+2D_vD_w\cos\left(\frac{4\pi n}{3}\right).$$
Because $F_{jk,n}=0$ if $n\neq 0$, $n(\mod 3)=0$ and $\cos(2\pi n/3)=\cos(4\pi n/3)=-1/2$ if  $n (\mod 3)=1,2$,
we find that
$$\langle D_{21}\tilde{z}_1^{(1)}\rangle=\frac{1}{2}\sum_{n\neq 0}\left[\frac{cT^2}{(2\pi)^2}\frac{|\hat{F}_{21,n}|^2}{n^2}-\frac{iT}{2\pi n}\hat{F}_{21,n}\hat{F}_{11,n}^{*}\right][(D_u-D_v)^2+(D_u-D_w)^2+(D_v-D_w)^2]=$$
$$\sum_{n=1}^{\infty}\left[\frac{cT^2}{(2\pi)^2}\frac{|\hat{F}_{21,n}|^2}{n^2}+\frac{T\mbox{Im}(\hat{F}_{21,n}\hat{F}_{11,n}^{*})}{2\pi n}\right][(D_u-D_v)^2+(D_u-D_w)^2+(D_v-D_w)^2].$$
Similarly,
$$\langle \tilde{D}_{22}\tilde{z}_2^{(1)}\rangle=\sum_{n\neq 0}\left(\frac{-iT}{4\pi n}\right)\hat{F}_{22,n}\hat{F}_{21,n}^*[(D_u-D_v)^2+(D_u-D_w)^2+(D_v-D_w)^2]=$$
$$\sum_{n=1}^{\infty}\frac{T}{2\pi n}\mbox{Im}(\hat{F}_{22,n}\hat{F}_{21,n}^*)[(D_u-D_v)^2+(D_u-D_w)^2+(D_v-D_w)^2]$$
and
$$\langle D_{23}\tilde{z}_3^{(1)}\rangle=\frac{1}{2}\sum_{n\neq 0}\frac{\hat{F}_{23,n}\hat{F}_{31,n}^{*}}{1-2n\pi i/T}[(D_u-D_v)^2+(D_u-D_w)^2+(D_v-D_w)^2]=$$
$$\sum_{n=1}^{\infty}\frac{\mbox{Re}(\hat{F}_{23,n}\hat{F}_{31,n}^{*})-(2\pi n/T)\mbox{Im}(\hat{F}_{23,n}\hat{F}_{31,n}^{*})}{1+(2n\pi/T)^2}[(D_u-D_v)^2+(D_u-D_w)^2+(D_v-D_w)^2].$$

\noindent Thus, we find that the dependence of the parameter $R$, which determines the splitting of the double eigenvalue, on the diffusion coefficients has a universal form
$$R=r[(D_u-D_v)^2+(D_u-D_w)^2+(D_v-D_w)^2],
$$
where the coefficient 
$$r=\sum_{n=1}^{\infty}\left[\frac{c^2T^2}{(2\pi)^2}\frac{|\hat{F}_{21,n}|^2}{n^2}+\frac{cT\mbox{Im}(\hat{F}_{21,n}\hat{F}_{11,n}^{*})}{2\pi n}+\frac{cT\mbox{Im}(\hat{F}_{22,n}\hat{F}_{21,n}^{*})}{2\pi n}+\right.$$
\begin{equation}
\left.+c\frac{\mbox{Re}(\hat{F}_{23,n}\hat{F}_{31,n}^{*})-(2\pi n/T)\mbox{Im}(\hat{F}_{23,n}\hat{F}_{31,n}^{*})}{1+(2n\pi/T)^2}\right]
\label{eq85}
\end{equation}
is determined solely by the solution of the undisturbed problem and does not depend on the diffusion coefficients.

Returning to expression (\ref{eq75}) for eigenvalues, we find 
\begin{equation}
\lambda_{\pm}^{(1)}=-\frac{1}{3}(D_u+D_v+D_w)\pm\sqrt{r[(D_u-D_v)^2+(D_u-D_w)^2+(D_v-D_w)^2]}.
\label{eq86}
\end{equation}
Thus, if $r\leq 0$, the long-wave spatial modulations of uniform oscillations decay. If $r>0$, the instability takes place if
$$(D_u+D_v+D_w)^2<\frac{\alpha}{2}[(D_u-D_v)^2+(D_u-D_w)^2+(D_v-D_w)^2],\;\alpha=18r,$$
or
\begin{equation}
\label{eq87}
(1-\alpha)(D_u^2+D_v^2+D_w^2)+(2+\alpha)(D_uD_v+D_uD_w+D_vD_w)<0
\end{equation}
For $\alpha<1$, the uniform oscillations are stable for any diffusion coefficients. If one of the diffusion coefficients, e.g., $D_u$ is not equal to zero (as we noted in Section \ref{sec:3}, it can be chosen equal to 1), while $D_v=D_w=0$, then the instability takes place for any $\alpha>1$. In the case $D_u=D_v=D_w$, the uniform oscillations are always stable.
\noindent For small amplitudes of oscillations, the higher Fourier components are
small, therefore it is sufficient to take into account just a few lowest
components. The basic solution of (\ref{eq12}) can be written as
$$u_0(t;a)=\frac{1}{3}+ae^{i\omega(|a|)t}+a^*e^{-i\omega(|a|)t}+a^2e^{2i\omega(|a|)t}+
a^{*2}e^{-2i\omega(|a|)t}+O(a^3),$$
$$v_0(t;a)=\frac{1}{3}+
a\left(-\frac{1}{2}-i\frac{\sqrt{3}}{2}\right)e^{i\omega(|a|)t}+
a^{*2}\left(-\frac{1}{2}+i\frac{\sqrt{3}}{2}\right)e^{-i\omega(|a|)t}+$$
\begin{equation}
a^2\left(-\frac{1}{2}+i\frac{\sqrt{3}}{2}\right)e^{2i\omega(|a|)t}+
a^{*2}\left(-\frac{1}{2}-i\frac{\sqrt{3}}{2}\right)e^{-2i\omega(|a|)t}+O(a^3),
\label{eq88}
\end{equation}
$$w_0(t;a)=\frac{1}{3}+
a\left(-\frac{1}{2}+i\frac{\sqrt{3}}{2}\right)e^{i\omega(|a|)t}+
a^*\left(-\frac{1}{2}-i\frac{\sqrt{3}}{2}\right)e^{-i\omega(|a|)t}+$$
$$a^2\left(-\frac{1}{2}-i\frac{\sqrt{3}}{2}\right)e^{2i\omega(|a|)t}+
a^{*2}\left(-\frac{1}{2}+i\frac{\sqrt{3}}{2}\right)e^{-2i\omega(|a|)t}+O(a^3),$$
where
$$\omega(|a|)=\omega_0+\omega_2|a|^2+O(|a|^4),\;\omega_0=\frac{\sqrt{3}}{3},\;
\omega_2=-2\sqrt{3}.$$
The relation between $a$ and $A$ can be found from (\ref{eq11}):
$$A=u_0v_0w_0=\frac{1}{27}-|a|^2+O(|a|^4),$$
hence $|a|^2=1/27-A$.

\noindent The calculation of the parameter $\alpha$ presented in Appendix B gives
\begin{equation}
\alpha=-9|a|^2+O(|a|^4).
\label{eq89}
\end{equation}
Thus, the small-amplitude oscillations are stable with respect to spatial modulations.

\begin{figure}
\centering
\includegraphics[width=0.6\linewidth]{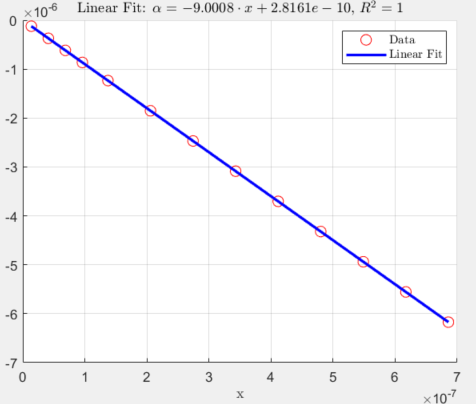}
\caption{\label{figure4} Plot of the stability parameter \(\alpha\) as a function of $x=1/27-A$.}
\end{figure}

\begin{figure}
\centering
\includegraphics[width=0.6\linewidth]{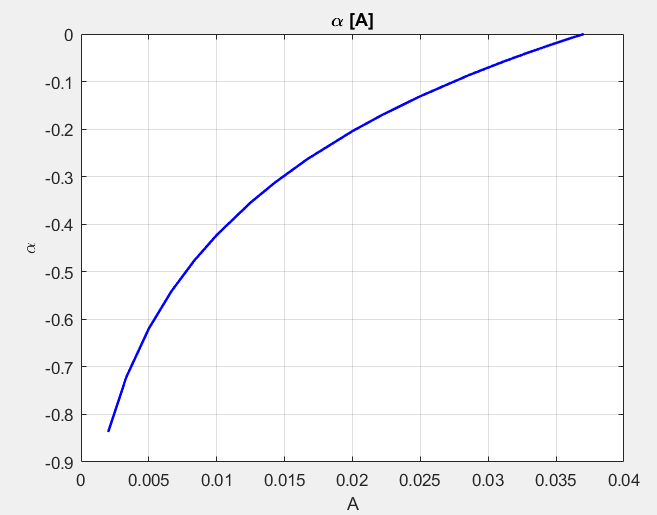}
\caption{\label{figure5} Dependence} of the stability parameter \(\alpha\)  on \(A\).
\end{figure}

\noindent For arbitrary $A$, the coefficient $\alpha$ was calculated numerically. Technically, it is more convenient to calculate the monodromy matrix rather than the matrix $B$. The results of the calculation are presented in Figures \ref{figure4}, \ref{figure5}. Figure \ref{figure4} confirms formula (\ref{eq89}): the linear fit of the dependence between  $\alpha$ and $x=1/27-A$ gives 
\begin{equation}
\label{eq90}
\alpha=-9.0008x+2.8161 \cdot 10^{-10} \ , \ R^2=1
\end{equation}
where \(R^2\) is the regression coefficient.

\noindent Computations carried out at a finite value of $x$ show that $\alpha(A)<0$ for arbitrary $A$ (see Figure \ref{figure5}). Hence, the uniform oscillations are stable with respect to long-wave modulations.
\section{\texorpdfstring{Stability analysis for arbitrary $k$}{Stability analysis for arbitrary k}}
\label{sec:5}
\subsection{\texorpdfstring{The case \(D_u=D_v=D_w\)}{The case Du=Dv=Dw}}
We saw in the previous section that in the case $D_u=D_v=D_w=D$ there is no splitting of the degenerate eigenvalue, and $\lambda\sim -Dk^2$ at small $k$. Actually, that is correct for any $k$.
\noindent Indeed,
assume that \(D_u=D_v=D_w=D>0\), so that system (\ref{eq22}) is
\begin{equation}
\label{eq91}
\dot{U}=-(Dk^2+1-2w_0)U-2u_0V,\;
\dot{V}=-(Dk^2+1-2u_0)V-2v_0W,\;
 \dot{W}=-(Dk^2+1-2v_0)W-2w_0U.
\end{equation}
Then the parameter $k$ can be eliminated by
changing variables: \\
\((U,V,W)=(\tilde{U},\tilde{V},\tilde{W})e^{-Dk^2t}\); 
\((\dot{U},\dot{V},\dot{W})=(\dot{\Tilde{U}}-Dk^2\Tilde{U},\dot{\Tilde{V}}-Dk^2\Tilde{V},\dot{\Tilde{W}}-Dk^2\Tilde{W})e^{-Dk^2t} \).  \\
The obtained system is equivalent to that at $k=0$:
\begin{equation}
\label{eq92}
\dot{\tilde{U}}=-(1-2w_0)\tilde{U}-2u_0\tilde{V},\;
\dot{\tilde{V}}=-(1-2u_0)\tilde{V}-2v_0\tilde{W},\;
\dot{\tilde{W}}=-(1-2v_0)\tilde{W}-2w_0\tilde{U}.
\end{equation} 
Thus, 
the fundamental matrix of system (\ref{eq91}) is 
$$\tilde{\Phi}(t)=e^{-Dk^2t}\Phi(t),$$
and the monodromy matrix of this system is
$$\Tilde{M}=e^{-Dk^2T}M,$$
 where $M$ is the monodromy matrix of the system with \(k=0\).  Hence, $\lambda(k)=\lambda(0)-Dk^2$ for all the 
eigenfunctions.
Therefore, in the case of \(D_u=D_v=D_w=D>0\) the spatially uniform oscillations are linearly stable with respect to disturbances with arbitrary $k$.
\subsection{\texorpdfstring{The case $D_u=1$, $D_v=D_w=0$}{The case Du=1, Dv=Dw=0}}
\begin{figure}[htbp]
\centering
\includegraphics[width=0.8\linewidth]{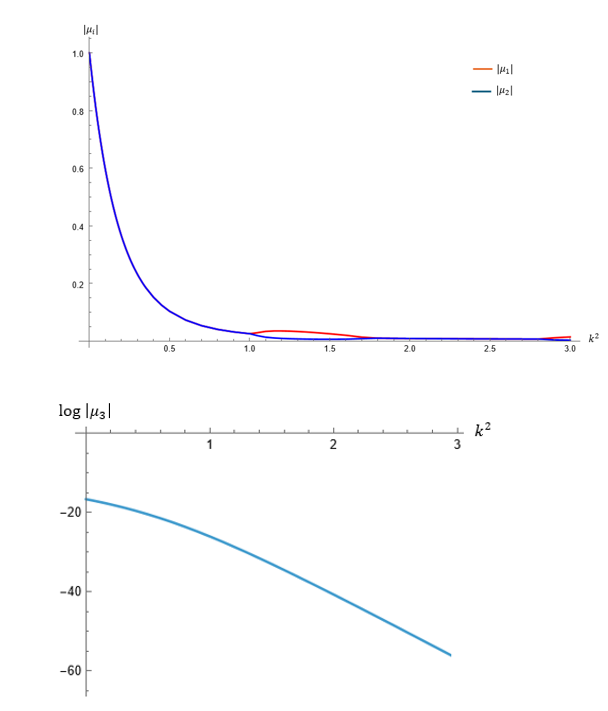}
\caption{\label{figure6} The moduli of the two large eigenvalues and the natural logarithm} of the small eigenvalue of the monodromy matrix for $T=16.70$, \(D_u=1,D_v=D_w=0\).\\
\end{figure}

\begin{figure}[htbp]
\centering
\includegraphics[width=1\linewidth]{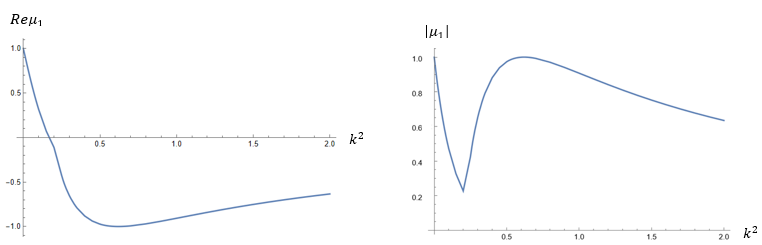}
\caption{\label{figure7} Modulus and real part of the leading multiplier} as a function of \(k^2\) for $T=T_*=22.86$, \(D_u=1,D_v=D_w=0\). The critical value $k_*^2\approx 0.60$.
\end{figure}

In Section \ref{sec:4}, we have seen that splitting of the eigenvalues at small $k$ is especially large in the case $D_u=1$, $D_v=D_w=0$ (see formula (\ref{eq86})), but that splitting affects only the imaginary part of the eigenvalue and hence does not influence the stability. Now we consider that case for arbitrary $k$ numerically. It will be shown that in this case a finite-wavenumber, period-doubling instability can occur. We calculate three linearly independent solutions of (\ref{eq22}) using exact solutions (\ref{eq16}), (\ref{eq17}) for the basic spatially uniform oscillations with a period $T$. Equations (\ref{eq22}) are integrated numerically during the time interval $0<t<T$, and the monodromy matrix $M(k^2)$ is found. The eigenvalues of that matrix (the multipliers) $\mu_i(k^2)$, $i=1,2,3$, determine the stability of the periodic solution (\ref{eq16}), (\ref{eq17}): if all $|\mu_i(k^2)|<1$ for any $k^2$, it is stable, and if $|\mu_i(k^2)|>1$ for a certain $i$ and $k^2$, it is unstable.
\noindent From the results of the previous sections we know that
for \(k=0\) the eigenvalues of matrix $B$ are $0$, $0$, and $-1$, hence the eigenvalues of the monodromy matrix are \(1,1,e^{-T}\) and the periodic solutions are orbitally stable (but not asymptotically stable, because one of the solutions of the linearized problem oscillates with linearly growing amplitude). 
\noindent At small $k$, according to (\ref{eq86}),
$$|\mu_{\pm}(k^2)|=|\exp(\lambda_{\pm}(k^2)T)|\sim\exp\left[-\frac{1}{3}(D_u+D_v+D_w)k^2T\right]<1,$$
because $r<0$, while $\mu_3(k^2)\sim\exp(-T)$, hence the solution is stable with respect to long-wave modulations.
\noindent The behavior of $\mu_i(k^2)$ at a finite value of $k$ depends on the value of the parameter $A$. It is more convenient to parameterize solutions (\ref{eq16}),(\ref{eq17}) by their period which is a monotonically decreasing function of $A$. A typical dependence of multipliers on $k^2$ for moderate values of $T$ is shown in Figure \ref{figure6}.
The multiplier $\mu_3$ is always real and is rather small. The multipliers $\mu_1$ and $\mu_2$ are complex with $\mu_2=\mu_1^*$ and $|\mu_1|=|\mu_2|<1$ in two intervals $0<k^2<k_{1}^2$ and $k_2^2<k^2<k_{3}^2$. In the intervals $k_{1}^2<k^2<k_2^2$ and $k^2>k_3^2$, $\mu_1$ and $\mu_2$ are real and non-equal. The modulus of the largest multiplier has a maximum at a certain value of $k^2$. This maximum grows with $T$. There exists a value $T_*$ such that for $T<T_*$ that maximum is lower than 1, so the disturbances decay with time. At $T=T_*$, that maximum becomes equal to 1 for a certain value $k^2=k_*^2$ (see Figure \ref{figure7}).
Actually, at this point the multiplier $\mu_1(k^2)=-1$ (see Figure \ref{figure7}).
Similarly, there exists a critical value \(A_*=5*10^{-4}\) such that for \(A>A_*\) the maximum is less than 1, and for \(A<A_*\) the maximum is greater than 1, and the maximum is decreasing with \(A\).
The corresponding solution of the linearized problem $x(t)=(U(t),V(t),W(t))$ is periodic with period $2T$, meaning that a period-doubling bifurcation occurred. Indeed, $x(t+T)=-x(t)$, hence $x(t+2T)=x(t)$.

\begin{figure}[htbp]
\centering
\includegraphics[width=0.6\linewidth]{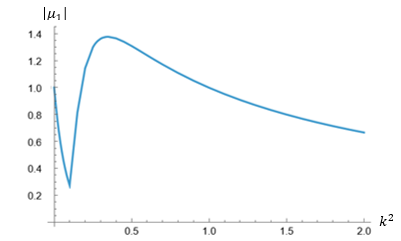}
\caption{\label{figure8} Modulus of the leading multiplier} as a function of \(k^2\) T=39.22, \(D_u=1,D_v=D_w=0\)
\end{figure}

\noindent For $T>T_*$, $|\mu_1(k^2)|>1$ in a certain interval $k_l^2<k^2<k_r^2$ (see Figure \ref{figure8}).
\newpage
\noindent At higher values of $k^2$, $\mu_1(k^2)$ can be empirically approximated as $\mu_1(k^2)\sim -2/(1+k^2)$.

\noindent Thus, spatially uniform oscillations with a sufficiently large period are unstable with respect to spatially non-uniform oscillations with the property
$$x(t+T)=-|\mu_1|x(t),\;x(t+2T)=|\mu_1|^2x(t).$$
\noindent The instability described above is similar to that found in \cite{ACR} for periodic solutions close to a homoclinic trajectory. In our case, the instability appears when with the growth of parameter $T$ the periodic solution becomes close to a heteroclinic cycle.
\subsection{Dependence of stability on diffusion coefficients}
\begin{figure}[htbp]
\centering
\includegraphics[width=0.7\linewidth]{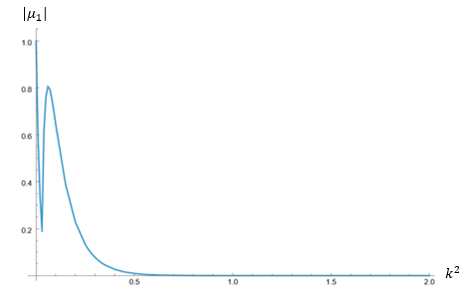}
\caption{\label{figure9} Modulus of the leading multiplier as a function of \(k^2\) for,
\(D_u=1,D_v=D_w=0.07\). There is no
instability even for T=142.84.}
\end{figure}

With the fixed value $D_u=1$ and an
increase of other diffusion coefficients, the critical period \(T_*\) grows fast, which means that 
the instability  appears only for periodic solutions that are closer to the heteroclinic cycle. An example of stabilization at nonzero $D_v$, $D_w$ is shown in Fig \ref{figure9}. 
In the case \(D_v=0,D_w=0.07\) we  have found the
instability.  We were unable to find an instability in the case \(D_v=0.07,D_w=0\).

\subsection{Generalization of the problem}
The spatially extended May-Leonard system (\ref{eq20}) belongs to a wider class of systems,
\begin{subequations}
\begin{equation}
    u_t=u(1-u-\alpha_1 v-\beta_1 w)+D_uu_{xx}
   \label{eq93a}
\end{equation}
\begin{equation}
   v_t=v(1-v-\beta_2 u-\alpha_2 w)+D_vv_{xx} 
   \label{eq93b}
\end{equation}
\begin{equation}
   w_t=w(1-w-\alpha_3 u- \beta_3 v)+D_ww_{xx}.
   \label{eq93c}
\end{equation}
\label{eq93}
\end{subequations}
The spatially homogeneous solutions of system (\ref{eq93}) have a property similar to that of the symmetric May-Leonard system: in the case where 

{\begin{equation}
\label{eq94}
(\alpha_1-1)(\alpha_2-1)(\alpha_3-1)=(1-\beta_1)(1-\beta_2)(1-\beta_3) \ , \ 0\leq\beta_i<1<\alpha_i
\end{equation}
the system has an attracting two-dimensional manifold containing a continuum of periodic orbits, and has no periodic orbits otherwise \cite{Chi} (with different notations). The approaches for studying the instabilities of those periodic solutions, which are described in preceding sections of the paper, can be applied to system (\ref{eq93}). Similarly, the trajectories close to the heteroclinic cycle, can be subject to a period-doubling instability with respect to spatially periodic disturbances. An example is shown in Figure \ref{figure10}.
\begin{figure}[htbp]
\centering
\includegraphics[width=0.7\linewidth]{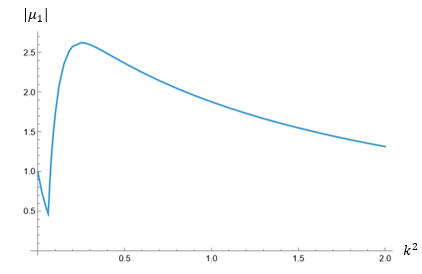}
\caption{\label{figure10} Modulus of the leading  multiplier} as a function of \(k^2\) for the parameters \(\alpha_1=10,\alpha_2=4,\alpha_3=28/27,\beta_1=\beta_2=\beta_3=0\); $T=96.07$, \(D_u=1,D_v=D_w=0\).There is an interval of wave-numbers where $|\mu_1|>1$, which corresponds to instability.
\end{figure}

\section{Conclusions}
\label{sec:6}
In this paper, the stability of periodic solutions of the extended symmetric May-Leonard system has been investigated.  Exact solutions  describing spatially uniform oscillations are found, and their linear stability with respect to spatially periodic disturbances is explored. A perturbative approach for studying the stability with respect to long-wave disturbances has been suggested and applied. It is shown that the periodic solutions are stable with respect to long-wave spatially periodic modulation. The spatially uniform solutions with sufficiently large temporal periods can be unstable with respect to spatially periodic disturbances with double temporal period. Possible generalizations of the problem are discussed.

 The authors are grateful to Michael Zaks for valuable discussions and his help in performing numerical computations.

\appendix
\section{\texorpdfstring{Fundamental matrix at $k=0$}{Fundamental matrix at k=0}}
In the case of \(k=0\)  (no dependence of $x$), we obtain the system
\begin{equation}
\dot{U}=-(1-2w_0)U-2u_0V,\;\dot{V}=-(1-2u_0)V-2v_0W,\;\dot{W}=-(1-2v_0)W-2w_0U.
\label{eq95}
\end{equation}
Equations (\ref{eq5}) and (\ref{eq10}) are valid, but they have to be applied in the linearized form as
\begin{equation}
U+V+W=C_1e^{-t}
\label{eq96}
\end{equation}
and
\begin{equation}
Uv_0w_0+u_0Vw_0+u_0v_0W-3A(U+V+W)=C_2,
\label{eq97}
\end{equation}
where $C_1$ and $C_2$ are some constants determined by the initial conditions (relations $u_0+v_0+w_0=1$ and $u_0v_0w_0=A$ are used in (\ref{eq97}).

\noindent The relations (\ref{eq96}) and (\ref{eq97}) make it possible to find analytical expressions in the form of integrals for all the linearly independent solutions of the system. First, let us consider the temporal evolution of disturbances within the invariant plane $u+v+w=1$, i.e., 
\begin{equation}
U+V+W=0
\label{eq98}
\end{equation}
$(C_1=0)$. Substituting $w_0=1-u_0-v_0$ and $W=-U-V$ into (\ref{eq95}) and (\ref{eq97}), we find:
\begin{subequations}
\label{eq99}
\begin{equation}
\dot{U} = U[1-2u_0-2v_0]-2u_0V \label{eq99a}
\end{equation}
\begin{equation}
\dot{V} = V[-1+2u_0+2v_0]+2v_0U \label{eq99b}
\end{equation}
\end{subequations}
and
\begin{equation}
v_0U(1-2u_0-v_0)+u_0V(1-u_0-2v_0)=C_2.
\label{eq100}
\end{equation}
Using equations (\ref{eq12}) written for $(u_0,v_0,w_0)$, we  can rewrite (\ref{eq100}) as
\begin{equation}
-\dot{v}_{0}U+\dot{u}_0V=C_2.
\label{eq101}
\end{equation}
From (\ref{eq101}) we find the dependence between U,V:
\begin{equation}
V=\frac{C_2+\dot{v_0}U}{\dot{u_0}}
\label{eq102}
\end{equation}
Substituting (\ref{eq102}) into (\ref{eq99a}) and using the relation
$$\ddot{u}_0=(1-2u_0-2v_0)\dot{u}_0-2u_0\dot{v}_0$$
obtained by the differentiation of (\ref{eq12}), we find:
\begin{equation}
\dot{u}_0\dot{U}-\ddot{u}_0U=-2C_2u_0.
\label{eq103}
\end{equation}

\noindent The homogeneous solution of (\ref{eq103}) gives the first solution of (\ref{eq95}),
\begin{equation}
U_1(t)=\dot{u}_0 \ , \ V_1(t)=\frac{\dot{v}_0U_1}{\dot{u}_0}=\dot{v}_0 \ , \ W_1(t)=-U_1-V_1=\dot{w}_0,
\label{eq104}
\end{equation}
which corresponds to the infinitesimal shift along the closed trajectory. This solution is periodic with period $T$.

Using the variation of parameter for finding the  solution of the non-homogeneous equation ($C_2\neq 0$), we obtain the second solution:
\begin{equation}
U_2(t)= -2C_2\dot{u_0} \int_0^t \frac{u_0(\tau)}{\dot{u}_0(\tau)^2} d\tau \ , \ V_2= \frac{C_2+\dot{v_0}U_2}{\dot{u_0}} \ , \ W_2=-U_2-V_2.
\label{eq105}
\end{equation}
Later on, we choose $C_2=-1/2$.

This disturbance corresponds to the motion on the invariant manifold $u+v+w=1$ along the close orbit with slightly changed $A$. Note that the second solution is not periodic: because the period $T(A)$ of the periodic solution (\ref{eq16}) depends on $A$, the difference between two solutions grows with time: $U_2(T)-U_2(0)=C_3\neq 0$.
\noindent It is convenient to decompose the function $U_2(t)$ into the periodic and non-periodic parts in the following way:
$$U_2(t)=U_{2p}(t)+c\dot{u}_0(t)t,$$
where 
$U_{2p}$ is $T$-periodic and $c=C_3/(\dot{u}_0(0)T)$.
Using (\ref{eq102}), we find that
$$V_2=V_{2p}+c\dot{v}_0(t)t.$$
Also, \(W_2=-U_2-V_2=-U_{2p}-V_{2p}-ct(\dot{u_0}+\dot{v_0})=W_{2p}+c\dot{w_0}t\), so  that:
\begin{equation}
U_2=U_{2p}+c\dot{u_0}t \ , \ V_2=V_{2p}+c\dot{v_0}t \ , \ W_2=W_{2p}+c\dot{w_0}t
\label{eq106}
\end{equation}
One can show that $V_{2p}(t)=U_{2p}\left(t-\frac{T}{3}\right)$ and $W_{2p}(t)=U_{2p}\left(t+\frac{T}{3}\right)$.
Obviously, the first solution and the second solution are linearly independent.
\noindent Note that we can get integral formulas similar to that for $U_2(t)$ in (\ref{eq105}) also for $V_2(t)$ and $W_2(t)$. In the numerical computations, we can apply different formulas at different intervals of $t$. 
\noindent In order to find the third solution, we use equations (\ref{eq96}) and (\ref{eq97}) with $C_1\neq 0$, $C_2=0$. Similarly to (\ref{eq101}), (\ref{eq102}) and (\ref{eq103}), we obtain:
\begin{equation}
-\dot{v_0}U+\dot{u_0}V=C_2+C_1e^{-t}(3A-u_0v_0)
\label{eq107}
\end{equation}
\begin{equation}
V=\frac{\dot{v_0}U+C_1e^{-t}(3A-u_0v_0)}{\dot{u_0}},
\label{eq108}
\end{equation}
and
\begin{equation}
{\dot{u_0}}\dot{U}-\ddot{u_0}U=-2C_1e^{-t}u_0(3A-u_0v_0).
\label{eq109}
\end{equation}
Using the variation of parameters,  we get the  the third particular solution:
\begin{equation}
U_3(t)=-2C_1\dot{u_0}\int_0^t e^{-\tau}u_0(\tau)\frac{3A-u_0(\tau)v_0(\tau)}{\dot{u_0(\tau)}^2}d\tau=U_{3p}(t)e^{-t},
\label{eq110}
\end{equation}
where $U_{3p}(t)$ is a periodic function with the period $T$.
\noindent Similarly, we find
\begin{equation}
\label{eq111}
V_3(t)=
V_{3p}e^{-t},  
\end{equation}
\begin{equation}
W_3(t)
=W_{3p}e^{-t},  \label{48c}
\end{equation}
with $V_{3p}$ and $W_{3p}$ being periodic functions,
\begin{equation}
V_{3p}(t)=U_{3p}(t-\frac{T}{3}) \ , \ W_{3p}(t)=
U_{3p}(t+\frac{T}{3})
\label{eq113}
\end{equation}
\noindent Summarizing the results obtained above, we find that 
the fundamental matrix for \(k=0\) is:
\begin{equation}
Y(t;0) =
\begin{pmatrix}
\dot{u_0} & U_{2p}+c\dot{u_0}t & U_{3p}e^{-t} \\
\dot{v_0} & V_{2p}+c\dot{v_0}t & V_{3p}e^{-t} \\
\dot{w_0} & W_{2p}+c\dot{w_0}t & W_{3p}e^{-t}
\end{pmatrix}
\label{eq114}
\end{equation}
 where $U_{jp}$, $V_{jp}$, $W_{jp}$, $j=2,3$, are periodic functions.
 \section{Long-wave stability of small amplitude oscillations}

In this appendix, we chose the initial point for time $t${} in such a way that the amplitude $a$ in (\ref{eq88}) is real. Also, we denote
\begin{equation}
 \exp(i\omega(a)t)\equiv C(t).
\label{eq115}
\end{equation}
\subsection{\texorpdfstring{Fundamental matrix at $k=0$}{Fundamental matrix at k=0}}
For each value of $A$, the stationary solutions
$(u_0(t+t_0,a),v_0(t+t_0,a),w_0(t+t_0,a)$ form a two-parametric family.
Their derivatives with respect to $t_0$ and $a$ are stationary solutions
of system (\ref{eq22}).

\noindent The first solution can be found according to formulas (\ref{eq26}). It is
convenient to divide it by $\omega_0a$ and take
$$U_1(t)=iC(t)-iC^{-1}(t)+2iaC^2(t)-2iaC^{-2}(t)+O(a^2),$$
$$V_1(t)=\left(-\frac{i}{2}+\frac{\sqrt{3}}{2}\right)C(t)+
\left(\frac{i}{2}+\frac{\sqrt{3}}{2}\right)C^{-1}(t)+$$
\begin{equation}
(-i-\sqrt{3})AC^2(t)+
(i-\sqrt{3})AC^{-2}(t)+O(A^2)=U_1(t-T/3),
\label{eq116}
\end{equation}
$$W_1(t)=\left(-\frac{i}{2}-\frac{\sqrt{3}}{2}\right)C(t)+
\left(\frac{i}{2}-\frac{\sqrt{3}}{2}\right)C^{-1}(t)+$$
$$(-i+\sqrt{3})aC^2(t)+
(i+\sqrt{3})aC^{-2}(t)+O(a^2)=U_1(t+T/3).$$

\noindent The second solution can be calculated as $(\partial u_0/\partial a,\;
\partial v_0/\partial a,\; \partial w_0/\partial a)$; that gives
$$(U_2,V_2,W_2)=(U_{2p},V_{2p},W_{2p})+ct(U_1,V_1,W_1),$$
where functions
$$U_{2p}(t)=C(t)+C^{-1}(t)+2aC^2(t)+2aC^{-2}(t)+O(a^2),$$
$$V_{2p}(t)=\left(-\frac{1}{2}-i\frac{\sqrt{3}}{2}\right)C(t)+
\left(-\frac{1}{2}+i\frac{\sqrt{3}}{2}\right)C^{-1}(t)$$
$$+(-1+i\sqrt{3})aC^2(t)+
(-1-i\sqrt{3})aC^{-2}(t)+O(a^2)=$$
$$U_{2p}(t-T/3),$$
$$W_{2p}(t)=\left(-\frac{1}{2}+i\frac{\sqrt{3}}{2}\right)C(t)+
\left(-\frac{1}{2}-i\frac{\sqrt{3}}{2}\right)C^{-1}(t)$$
$$+(-1-i\sqrt{3})aC^2(t)+
(-1+i\sqrt{3})aC^{-2}(t)+O(a^2)=$$
$$\tilde{U}_2(t+T/3)$$
are periodic, and
$$c=2\omega_2a^2=-4\sqrt{3}a^2.$$

\noindent The third solution can be found by the ansatz
$$(U_3,V_3,W_3)=e^{-t}(U_{3p},V_{3p},W_{3p}),$$
where functions $U_{3p}$, $V_{3p}$, $W_{3p}$ are periodic.
We obtain:
$$U_{3p}(t)=1+(3-i\sqrt{3})aC(t)+(3+i\sqrt{3})aC^{-1}(t)+O(a^2),$$
$$V_{3p}(t)=1+(-3-i\sqrt{3})aC(t)+(-3+i\sqrt{3})aC^{-1}(t)+O(a^2)=
U_{3p}(t-T/3),$$
$$W_{3p}(t)=1+2i\sqrt{3}aC(t)+(-2i\sqrt{3})aC^{-1}(t)+O(a^2)=
U_{3p}(t+T/3).$$

The obtained fundamental matrix $\Phi(t)$ can be presented in the form
$$\Phi(t)=P(t)\exp(Bt),$$
where $P(t)$ and $B$ are determined by formulas (\ref{eq31}), (\ref{eq32}) with one
modification: the first column of $P$ is formed by functions (\ref{eq116})
rather than by $(\dot{u}_0,\dot{v}_0,\dot{w}_0)$.

\noindent The matrix $P(t)$ has the form (\ref{eq99b}) with $P_{11}=\tilde{U}_1$,
$P_{12}=\tilde{U}_2$ and $P_{13}=\tilde{U}_3$. For the inverse matrix we
find:
$$(P^{-1})_{11}=\frac{i}{6}C(t)-\frac{i}{6}C^{-1}(t)-\frac{i}{3}aC^2(t)+
\frac{i}{3}aC^{-2}(t)+\frac{a}{\sqrt{3}}+O(a^2),$$
$$(P^{-1})_{21}=\frac{1}{6}C(t)+\frac{1}{6}C^{-1}(t)-\frac{1}{3}aC^2(t)-
\frac{1}{3}aC^{-2}(t)-A+O(a^2),$$
$$(P^{-1})_{31}=\frac{1}{3}+O(a^2).$$

\subsection{Splitting of eigenvalues}

The elements of the matrix F(t) are:
$$F_{11}(t)=(P^{-1})_{11}P_{11}=-\frac{1}{6}C^{-2}-\frac{i}{\sqrt{3}}aC^{-1}+
\frac{1}{3}+\frac{i}{\sqrt{3}}aC-\frac{1}{6}C^2+O(a^2),$$
$$F_{12}(t)=(P^{-1})_{11}P_{12}=-\frac{i}{6}C^{-2}+
\left(\frac{1}{\sqrt{3}}+\frac{2}{3}i\right)aC^{-1}+
\left(\frac{1}{\sqrt{3}}-\frac{2}{3}i\right)aC+\frac{i}{6}C^2+O(a^2),$$
$$F_{13}(t)=(P^{-1})_{11}P_{13}=
\left(\frac{\sqrt{3}}{6}-\frac{i}{6}\right)aC^{-2}-\frac{i}{6}C^{-1}+
\frac{i}{6}C+\left(\frac{\sqrt{3}}{6}\right)aC^2+O(a^2),$$
$$F_{21}(t)=(P^{-1})_{21}P_{11}=-\frac{i}{6}C^{-2}+\frac{i}{3}AC^{-1}-
\frac{i}{3}aC+\frac{i}{6}C^2+O(a^2),$$
$$F_{22}(t)=(P^{-1})_{21}P_{12}=\frac{1}{6}C^{-2}-aC^{-1}+
\frac{1}{3}-aC+\frac{1}{6}C^2+O(a^2),$$
$$F_{23}(t)=(P^{-1})_{21}P_{13}=
\left(\frac{1}{6}+\frac{i\sqrt{3}}{6}\right)aC^{-2}+\frac{1}{6}C^{-1}
+\frac{1}{6}C+\left(\frac{1}{6}-\frac{i\sqrt{3}}{6}\right)aC^2+O(a^2),$$
$$F_{31}(t)=(P^{-1})_{31}P_{11}=-\frac{2i}{3}aC^{-2}-\frac{i}{3}aC^{-1}+
\frac{i}{3}aC+\frac{2i}{3}aC^2+O(a^2),$$
$$F_{32}(t)=(P^{-1})_{31}P_{12}=\frac{2}{3}aC^{-2}+\frac{1}{3}C^{-1}+
\frac{1}{3}C+\frac{2}{3}aC^2+O(a^2),$$
$$F_{33}(t)=(P^{-1})_{31}P_{13}=
\left(1+\frac{i\sqrt{3}}{3}\right)aC^{-1}+
\frac{1}{3}+\left(1-\frac{i\sqrt{3}}{3}\right)aC+O(a^2).$$

\noindent Substituting obtained expressions for $F_{jk}$ into expression (\ref{eq85}), we find
$$r=-\frac{1}{2}a^2$$
hence
$$\alpha=18r=-9a^2.$$

\end{document}